\documentclass[onecolumn,aps,prl,showpacs,superscriptaddress,notitlepage,nofootinbib]{revtex4-1}
\usepackage[a4paper,top=3cm,bottom=3cm,left=3.6cm,right=3.6cm]{geometry}
\usepackage[english]{babel}
\usepackage{graphicx}				
\usepackage{booktabs}			
\usepackage{amsmath}
\usepackage{amssymb}			
\usepackage{mathtools}
\usepackage{gensymb}				
\usepackage{braket}					
\usepackage{multirow}
\usepackage{setspace}
\usepackage[dvipsnames]{xcolor}
\usepackage[separate-uncertainty=true]{siunitx}

\makeatletter
\let\cat@comma@active\@empty
\makeatother

\begin{document}
\title{Creating mid-infrared single photons}

\author{Richard McCracken} 
\affiliation{Scottish Universities Physics Alliance (SUPA), Institute of Photonics and Quantum Sciences, School of Engineering and Physical Sciences, Heriot-Watt University, Edinburgh EH14 4AS, UK}
\author{Francesco Graffitti}
\affiliation{Scottish Universities Physics Alliance (SUPA), Institute of Photonics and Quantum Sciences, School of Engineering and Physical Sciences, Heriot-Watt University, Edinburgh EH14 4AS, UK}
\author{Alessandro Fedrizzi}
\affiliation{Scottish Universities Physics Alliance (SUPA), Institute of Photonics and Quantum Sciences, School of Engineering and Physical Sciences, Heriot-Watt University, Edinburgh EH14 4AS, UK}

\begin{abstract}
Single-photon creation through parametric downconversion underpins quantum technology for quantum sensing and imaging. Here we numerically study the creation of single photons in the near- and mid-infrared regime from $1.5$--$12~\mu$m in a range of novel nonlinear semiconductor and chalcopyrite materials. We identify phase-matching conditions and single out regimes in which group-velocity matching can be achieved with commercially available pump lasers. Finally, we discuss how mid-infrared single photons can be detected. Using our numerical results, we identify materials and pump lasers for up-conversion detection in conventional wavelength bands. Our study provides a complete recipe for mid-IR single-photon generation and detection, opening up quantum enhancements for mid-IR applications such as bio-medical imaging, communication, and remote sensing.
\end{abstract}

\maketitle

\section{Introduction}
Single-photon creation with quantum emitters or through parametric downconversion is a mature quantum technology with bright photon sources now routinely available at visible and telecommunications wavelengths. A major application area for single-photon sources is in quantum sensing and metrology, where they can exceed noise limitations intrinsic to classical systems for e.g. sub-shot-noise phase estimation \cite{slussarenko2017unconditional} or absorption measurements in the few-photon regime \cite{moreau2017demonstrating}. 

Expanding single-photon technology beyond the near-infrared ($1$ to $2$~$\mu$m) into the mid-infrared spectrum ($2$ to $20$~$\mu$m) could enable quantum advantages for a host of mid-IR applications~\cite{Ebrahim-Zadeh2007}, e.g. medical imaging \cite{Fernandez2005,Amrania2012} at ultra-low light levels. It would further open up access to scatter-free atmospheric windows \cite{Martini2002,Temporao2008,Bellei2016} for free-space quantum communication, and for quantum remote sensing, e.g. the detection of biological or chemical samples in the few-photon regime, stealth range finding, and in particular quantum LIDAR \cite{tan2008quantum,Wang2016}.  

The most popular method for generating single-photon pairs is parametric downconversion in crystals such as BBO, BiBO, PPLN and PPKTP. Due to their optical properties, they allow access from ultraviolet to near-infrared wavelengths. PPLN and PPKTP are suitable for generation of photons up to 5~$\mu$m, and single-photon generation and up-conversion detection in PPLN has been demonstrated for up to 4 $\mu$m \cite{Sua2017,Mancinelli2017}. However, beyond that limit these two materials become opaque. A suite of novel nonlinear materials has recently been explored using widely tunable optical parametric oscillators (OPOs). Four of the most exciting materials are OPGaP, OPGaAs, CSP and ZGP, each of which exhibit transparencies that extend well beyond 5~$\mu$m. 

Here we numerically study mid-infrared photon generation in these materials, benchmarking against known results in PPLN and PPKTP. We show that wavelengths up to 13~$\mu$m can comfortably reached with available pump lasers. We identify parameter regimes for type-0, type-I and type-II phase-matching and highlight a number of special cases for which group-velocity matching between the pump and signal/idler photons can be achieved, allowing for the creation of spectrally pure single photons without spectral filtering---a key requirement for applications requiring high photon-collection efficiency. Finally, we discuss how mid-infrared single photons can be detected, either with low band-gap semiconductor avalanche photo-detectors, superconducting nano-wire single-photon detectors, or frequency up-conversion to more conventional wavelengths.

\section{Parametric downconversion and group-velocity matching}

High-energy photons (commonly referred to as `pump' photons) passing through a nonlinear optical material can interact with the medium, decaying into two lower-energy photons (namely `signal' and `idler') under the conservation of energy and momentum. This three-wave mixing process is known as parametric downconversion (PDC) and is a widespread technique for generating high-quality single photons for optical quantum technologies. In fact, despite their probabilistic nature, recent schemes have demonstrated that PDC sources can approximate a nearly-deterministic photon source with high production rates and photon purity \cite{Broome2011,Collins2013,Xiong2016,Kaneda2017,Graffitti2018}

The spectral properties of the downconverted photons reflect the pump spectrum and the crystal nonlinear properties through the dispersion relations in the material. In particular, the PDC photons usually emerge from the crystal in a spectrally-correlated state that leads to poor heralded-photon spectral purities \cite{URen2006}: this is inconvenient for most of the photonics quantum applications that indeed require highly pure photons. For this reason, different techniques have been introduced for minimising the spectral correlations in the PDC photon pair. 

The simplest method is applying narrowband spectral filters to the photons. However, this comes at a significant cost in terms of optical loss, compromising the overall efficiency of the source.

A more sophisticated approach is tuning the experimental parameters to fulfil the group-velocity matching (GVM) condition \cite{Grice2001,URen2006,Mosley2008,Graffitti2017,Quesada2018}.
This method consists in choosing the group velocities of the three photons involved in the process so that the dispersion parameter $D = -(\text{GD}_p - \text{GD}_s)/(\text{GD}_p - \text{GD}_i)$ is greater than 0, where GD are the group delays of the photons (or equivalently, their inverse group velocities) \cite{Kaneda2016,Laudenbach2017}.
The dispersion parameter is equal to the $\text{tan}^{-1}\left(\theta\right)$, where $\theta$ is the angle between the phase-matching function and the x-axis in the signal-idler frequency space: for a full description of phase-matching function and the PDC spectral properties we refer the reader to \cite{URen2006,Laudenbach2017,Quesada2018}.

Whenever the GVM condition is satisfied, it is possible to maximise the heralded-photon purity by appropriately tuning pump spectral width respect to the crystal length \cite{URen2006,Quesada2018}.
When $D=0$ or $D=+\infty$~--~conditions known as asymmetric GVM, corresponding to $\theta=0$ and $\theta=90$, respectively~--~a spectrally-uncorrelated bi-photon state is asymptotically achieved for very spectrally-broad pump pulses. This condition provides high heralded-photon purities but it's not suitable for multiphoton experiments, as the signal and idler photons have different bandwidths and are therefore distinguishable.
$D=1$ (i.e. $\theta=45$) corresponds to symmetric GVM: the PDC photons produced under this condition are indistinguishable, but their spectral purity is limited by residual correlations rising from the uniform nonlinearity profile of the crystal \cite{Branczyk2011}.
In the case of poled crystals, it's possible to tailor the longitudinal nonlinearity of the crystal by changing its poling structure to further reduce the spectral correlations between the PDC photons
\cite{Branczyk2011,Dixon2013,Tambasco2016,Dosseva2016,Graffitti2017,Graffitti2018}.

For the purpose of this study we will identify the group-velocity matching region for a number of crystals generating in the near- and mid-IR, pointing out the distinctive regimes for which the dispersion parameter is either 0, 1 or $+\infty$.

\section{Mid-infrared nonlinear optical crystals}

Here we detail key optical properties of promising nonlinear crystals for mid-infrared single photon generation, including material transparency, crystallographic orientation, and the primary nonlinear tensor coefficient $(d_{jk})$ that can be exploited for efficient downconversion \cite{Boyd2008}. We also provide examples of mid-infrared generation in these materials, however we note that this survey is not exhaustive.

\subsubsection{Birefringent oxide crystals}

\begin{figure*}[htb]\center
\includegraphics[width=0.8\textwidth]{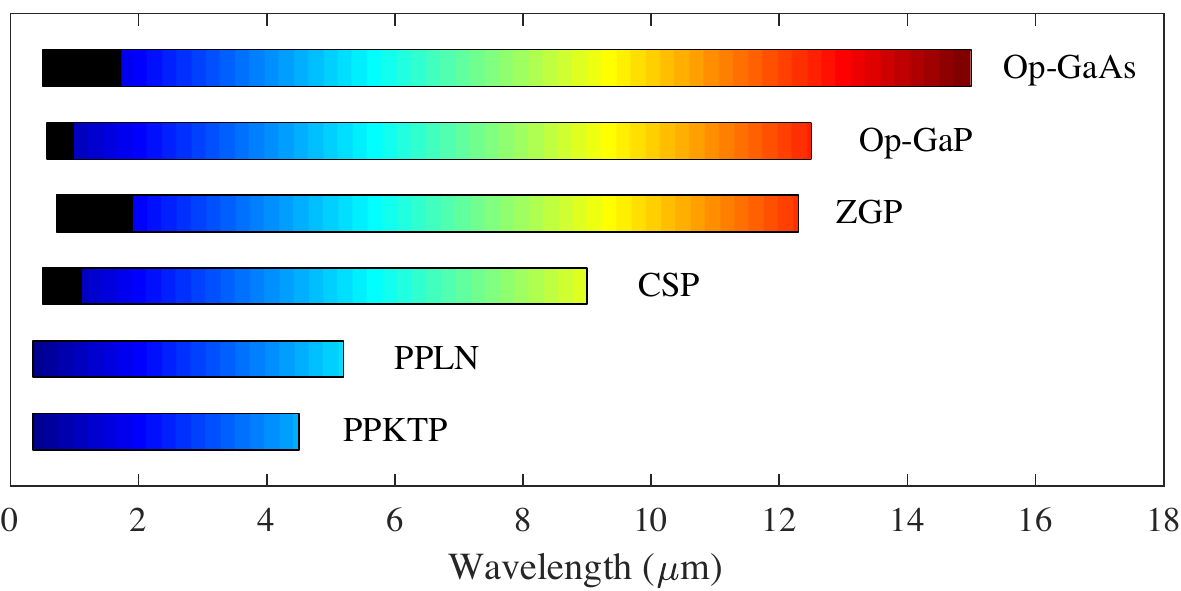}
\caption{Optical transparencies of several mid-infrared nonlinear crystals, with PPKTP and PPLN included for comparison. Areas shaded in black are transparent but display strong two-photon absorption.}
\label{fig1}
\end{figure*}

The most commonly employed nonlinear materials for parametric downconversion are lithium niobate (LiNbO$_3$, LN) and potassium titanyl phosphate (KTiOPO$_4$, KTP)~--~material properties and Sellmeier equations can be found in references \cite{Nikogosyan2005,Dmitriev1999a,Gayer2008b,Gayer2008,Zhao2010a,Fradkin1999d,Kato2002}. Non-critical quasi-phase-matching (QPM) in these materials typically requires the addition of a periodic structure, achieved by applying a lithographic electrode mask across the crystal, followed by electric field poling of periodic domains of alternating polarity. This poling technology is very mature, enabling highly tailored periodicities such as fan-out gratings, cascaded processes \cite{Balskus2015}, chirped structures \cite{tillman2003} and entirely domain-engineered crystals \cite{Kornaszewski2008a,Graffitti2017}. These oxide crystals are transparent across the entire visible and near-infrared spectrum and can be pumped by commercial Ti:sapphire (800~nm), Yb:fiber (1040~nm) and Er:fiber (1550~nm) laser sources, however their use in mid-infrared frequency conversion is limited due to the onset of multi-phonon absorption losses above 4.5~$\mu$m. In contrast, birefringent chalcopyrite crystals such as zinc germanium phosphide (ZnGeP$_2$, ZGP) and cadmium silicon phosphide (CdSiP$_2$, CSP), and semiconductors such as optically-patterned gallium phosphide (OP-GaP) and gallium arsenide (OP-GaAs) exhibit strong two-photon absorption at lower wavelengths but excellent transparency into the mid-infrared (Figure \ref{fig1}), along with significantly larger nonlinear coefficients than either LN or KTP (Table \ref{tableDeff}). 

\subsubsection{Birefringent chalcopyrite crystals}

Zinc germanium phosphide (ZGP) is a positive uniaxial crystal (\textit{n}$_e$ $>$ \textit{n}$_o$) with d$_{36}$ = 75.4~pmV$^{-1}$, enabling type-I ($o \to e + e$) and type-II ($o \to e + o$) birefringent phase-matching \cite{Boyd1972,Schunemann1998,Zelmon2001}. The crystal is transparent from 0.7~--~12.3~$\mu$m however strong two-photon absorption prevents pumping below 1.9~$\mu$m, therefore ZGP OPOs are typically pumped by master-oscillator-power-amplifier systems comprising a thulium (Tm)-based pump and a holmium (Ho)-based amplifier \cite{Hemming2013, Wang2017}, or with Tm-pumped Cr$^{3+}$:ZnSe lasers \cite{McKinnie2002}. Alternative pumping schemes have employed a cascaded optical parametric oscillators architecture, utilising a primary OPO to pump a secondary ZGP OPO \cite{Ganikhanov2001, Henriksson2007}, however this can introduce additional noise into the system.\\

\begin{table}
\caption{\label{tableDeff}Effective nonlinearity coefficients in all polarization configurations for the six crystals discussed in this paper. Maximum values for \textit{d\textsubscript{eff}} were calculated using the software \textit{SNLO~v70}, developed by \textit{AS-Photonics, LLC} \cite{SNLO} - note that precise experimental values depend on wavelength, crystal orientation and doping concentration. The letter \textit{o} denotes polarization along the \textit{ordinary} (\textit{y}-) axis, while the letter \textit{e} denotes polarization along the \textit{extraordinary} (\textit{z}-) axis.}
\medskip
\setlength{\tabcolsep}{8pt}
\begin{tabular}{cccccccc}
\hline
\multicolumn{2}{c}{Phase-matching}  & \multicolumn{6}{c}{Maximum effective nonlinear $\lvert \textit{d\textsubscript{eff}}  \rvert$ [pmV\textsuperscript{-1}] }\\
\multicolumn{2}{c}{type}  & PPKTP  & PPLN & OPGaP & OPGaAs & CSP & ZGP\\
\hline
\multirow{2}{*}{0} & $o \to o + o$ & 0 & 0 & 0 & 0 & N/A & N/A \\     
& $e \to e + e$ & 15.3 & 25 & 75 & 95 & N/A & N/A \\    
\hline     
\multirow{2}{*}{I} & $o \to e + e$ & 0 & 0 & N/A & N/A & 0 & 75.4 \\     
& $e \to o + o$ & 3.9 & 4.6 & N/A & N/A & 84 & 0 \\    
\hline
\multirow{2}{*}{II} & $o \to e + o$ & 3.9 & 4.6 & N/A & N/A & 0 & 75.4 \\     
& $e \to o + e$ & 0 & 0 & N/A & N/A & 84 & 0 \\   
\hline          
\end{tabular}
\end{table}

Cadmium silicon phosphide (CSP) is a negative uniaxial crystal (\textit{n}$_o$ $>$ \textit{n}$_e$) with d$_{36}$ = 84 pmV$^{-1}$, enabling type-I ($e \to o + o$) and type-II ($e \to o + e$) birefringent phase-matching \cite{Kemlin2011,Wei2018a}. The crystal is transparent from 0.5 - 9.0 $\mu$m, however the band edge is close to 500~nm, implying that two-photon absorption will  be present below 1.0~$\mu$m. Pulse-energy-dependent two-photon effects have been observed using ytterbium (Yb)-doped solid-state sources at 1.064 and 1.053~$\mu$m, suggesting that power scaling may require pumping with wavelength-shifted Yb:fiber amplifiers, or moving to 1.5 $\mu$m erbium (Er)-based technology. CSP-based OPOs producing picosecond \cite{Peremans2009, Kumar2011, Kumar2018} and femtosecond \cite{Kumar2015, Zhang2013c} pulses in the 6 - 7~$\mu$m region have been demonstrated when pumped near 1~$\mu$m, and an architecture employing intracavity cascaded nonlinear crystals has extended this to 8.1~$\mu$m \cite{Ramaiah-Badarla2016}. While the majority of CSP-based OPOs have exploited non-critical phase-matching ($\theta$ = 90$\degree$), type-I critical phase-matching has recently been demonstrated \cite{ODonnell2018}.

\subsubsection{Quasi-phase-matched semiconductors}
Semiconductors offer numerous advantages over traditional oxide crystals, as they have excellent thermal conductivity, high nonlinear coefficients and can be grown with high purity. Orientation-patterned gallium arsenide (OP-GaAs) was the first quasi-phase-matched semiconductor material \cite{Eyres2000}, offering two-photon-free transparency from 1.73~$\mu$m to above 15~$\mu$m and a high nonlinearity of d$_{36}$~=~95~pmV$^{-1}$ \cite{Skauli2003}. As semiconductors do not posses the ferroelectric properties of oxide crystals, conventional poling methods cannot be employed. Instead, a combination of molecular beam epitaxy, photolithography and selective etching followed by high-growth-rate hydride vapour phase epitaxy produces a 'thick-film' ($>$1.5~mm) layer with good grating integrity \cite{Schunemann2016, Lynch2008}. Growth, processing and patterning of semiconductor materials has matured significantly in recent years, with commercial samples available and downconversion demonstrated in several embodiments. Spurred by interest in mid-infrared frequency combs, doubly-resonant OP-GaAs OPOs were extensively developed in Stanford, pumped by chromium (Cr):ZnSe ($\sim$2.4~$\mu$m \cite{Vodopyanov2011, Smolski2015}) or Tm:fiber ($\sim$2.0~$\mu$m \cite{Lee2013a, Leindecker2012a}). A singly-resonant (non-degenerate) OP-GaAs OPO pumped by an amplified Tm:fiber laser was demonstrated by Heckl et al\cite{Heckl2016}, and was tuneable from 3~--~6~$\mu$m.

Orientation-patterned gallium phosphide (OP-GaP) was specifically developed in order to have a nonlinear semiconductor material that could be pumped by mature Yb, neodymium (Nd) and Er-based sources. The two-photon absorption edge is shifted to 1~$\mu$m and the crystals are transparent to 12.5~$\mu$m, with d$_{36}$~=~75~pmV$^{-1}$ \cite{Wei2018}. The patterning process is similar to that of OP-GaAs, with $>$1~mm layer thicknesses achieved repeatably \cite{Tassev2013}. Parametric downconversion has been demonstrated using a variety of pump lasers, including Yb:fiber \cite{Maidment2016, Kara2017a, Sorokin2018}, Er:fiber \cite{Ru2017}, Q-switched Nd:YVO$_4$  \cite{Pomeranz2015} and Nd:YAG lasers \cite{Ye2017}, and through difference frequency generation \cite{Lee2017, Sotor2018}.

\section{Numerical studies for phase-matching and group-velocity matching}

We now perform phase-matching and group-velocity-matching calculations for collinear type-0, type-I and type-II interactions in the nonlinear materials described in the previous section, with calculations verified against published OPO results. Using a range of pump wavelengths, we numerically solve the phase-matching equation

\begin{equation}\notag
\Delta k = 2\pi\Bigg(\frac{n(\lambda_p)}{\lambda_p} - \frac{n(\lambda_s)}{\lambda_s} - \frac{n(\lambda_i)}{\lambda_i} - \frac{1}{\Lambda}\Bigg)
\end{equation}

\noindent{where $\Delta k$ is the wave-vector mismatch and $n(\lambda)$ is the wavelength-dependent refractive index for the pump, signal and idler photons, determined from the Sellmeier equations for each crystal. Where temperature-dependent Sellmeier equations are available, we perform our calculations at 300~K. Where multiple Sellmeier equations exist, an average is taken. For each wavelength we calculate the grating period $\Lambda$ (or equivalent birefringently phase-matched crystal length) required to achieve $\Delta k = 0$, maximising the phase-matching efficiency. Similarly, we calculate the dispersion parameter $D$ described previously, displaying our results as group-velocity mismatch angle $\theta$ as detailed in \cite{URen2006, Quesada2018, Kaneda2016, Laudenbach2017}. We restrict our analysis to pump wavelengths above the two-photon absorption limit, and idler wavelengths below the transparency cut-off.

\subsubsection{Type-0 downconversion}

Type-0 downconversion provides access to the highest nonlinearities for grating-based crystals, yielding increased photon-pair generation rates and enabling efficient system architectures. GVM angle $\theta$ and grating period $\Lambda$ are calculated for OP-GaP and OP-GaAs, using PPKTP and PPLN as illustrative comparisons, and results are displayed in Figures \ref{fig2} to \ref{fig5}. In the case of type-0 phase-matching it is important to note that degenerate (spectrally symmetric) downconversion does not provide high spectral purity or indistinguishability due to a singularity in $\theta$, however non-degenerate downconversion is readily achievable.

\begin{figure*}[htb]\center
\includegraphics[width=0.96\textwidth]{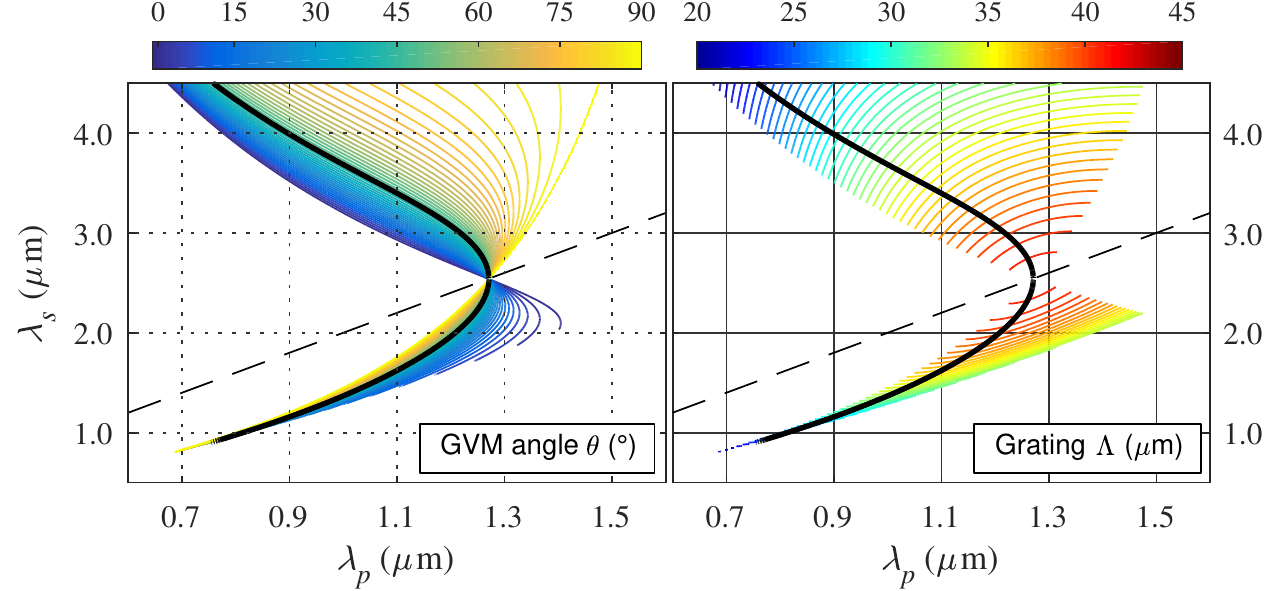}
\caption{Signal wavelength vs pump wavelength with corresponding GVM angle $\theta$ (left) and grating period $\Lambda$ (right) for type-0 ($e \to e + e$) downconversion in PPKTP. The dashed line indicates degeneracy, i.e. spectrally symmetric downconversion. The black line indicates wavelength configurations for high indistinguishability ($\theta = 45$), while the plotted boundaries indicate areas of high spectral purity ($\theta = 0, 90$).}
\label{fig2}
\end{figure*}

\begin{figure*}[htb]\center
\includegraphics[width=0.96\textwidth]{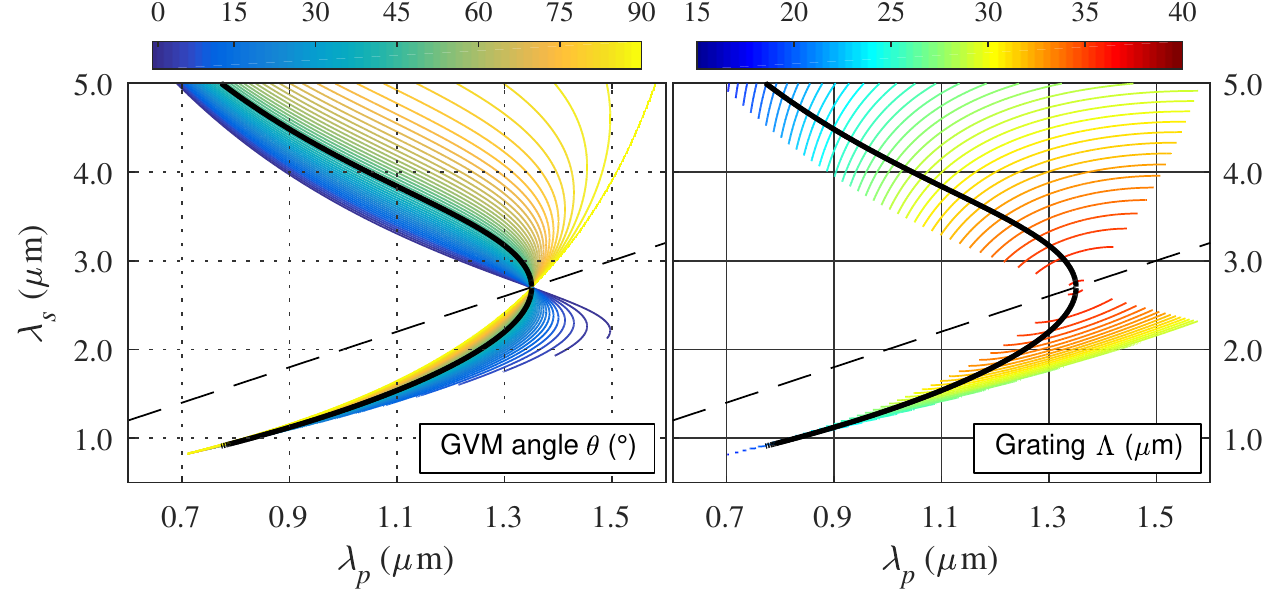}
\caption{Signal wavelength vs pump wavelength with corresponding GVM angle $\theta$ (left) and grating period $\Lambda$ (right) for type-0 ($e \to e + e$) downconversion in PPLN.}
\label{fig3}
\end{figure*}

\begin{figure*}[htb]\center
\includegraphics[width=0.96\textwidth]{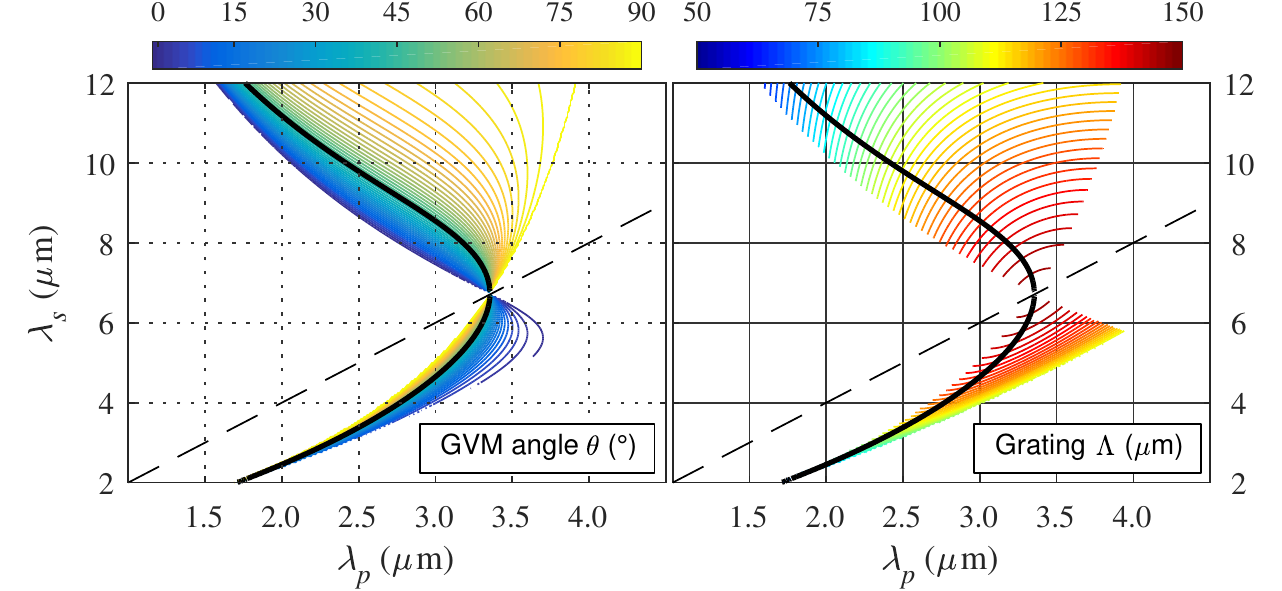}
\caption{Signal wavelength vs pump wavelength with corresponding GVM angle $\theta$ (left) and grating period $\Lambda$ (right) for type-0 ($e \to e + e$) downconversion in OP-GaP.}
\label{fig4}
\end{figure*}

\begin{figure*}[htb!]\center
\includegraphics[width=0.96\textwidth]{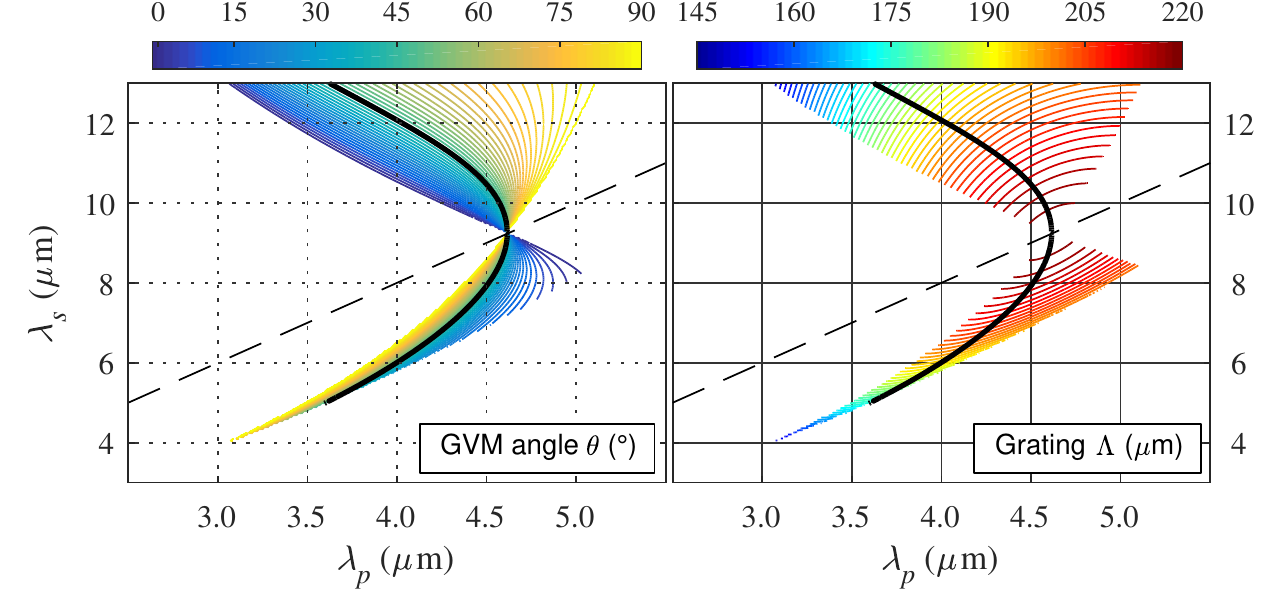}
\caption{Signal wavelength vs pump wavelength with corresponding GVM angle $\theta$ (left) and grating period $\Lambda$ (right) for type-0 ($e \to e + e$) downconversion in OP-GaAs.}
\label{fig5}
\end{figure*}

\subsubsection{Type-I downconversion}

Type-I downconversion provides parallel polarization states for the signal and idler photons, which are orthogonal to the pump polarization. This process is ideal for high-purity photons with low spectral correlation, however degenerate phase-matching is not achievable in bulk crystals due to a singularity in the GVM condition. Type-I degenerate downconversion has been achieved in waveguides in a backwards-propagating regime \cite{Christ2009a}, however the additional phase-matching terms introduced by waveguide dispersion are outside the scope of this investigation. Figure \ref{fig6} displays our calculations for PPLN ($e \to o + o$), in which we highlight cases of high spectral purity ($\theta~=~0,90$) and indistinguishability ($\theta~=~45$). While non-critical-phase-matching calculations for ZGP ($o \to e + e$) and CSP ($e \to o + o$) indicate regions of high purity, the required crystal lengths are \textless100~$\mu$m and are therefore unlikely to allow for generation of single photons at reasonable brightness. PPKTP type-I phase-matching does not permit the generation of spectrally pure heralded photons.\\

\begin{figure*}[htb]\center
\includegraphics[width=0.96\textwidth]{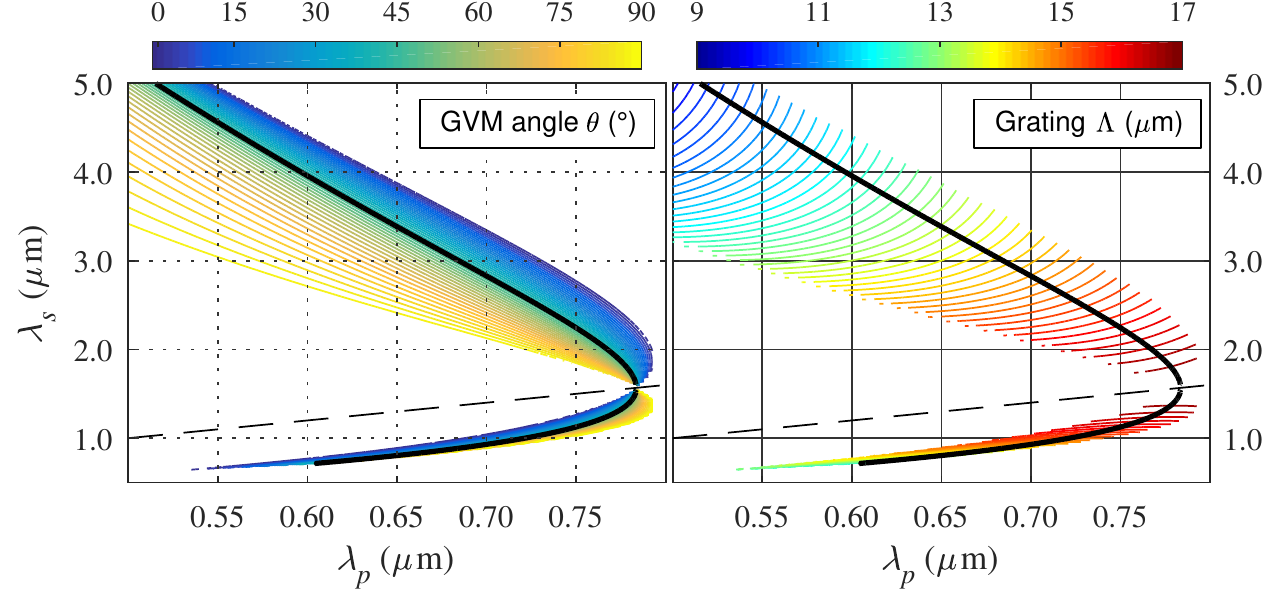}
\caption{Signal wavelength vs pump wavelength with corresponding GVM angle $\theta$ (left) and grating period $\Lambda$ (right) for type-I ($e \to o + o$) downconversion in PPLN.}
\label{fig6}
\end{figure*}

\subsubsection{Type-II downconversion}

Type-II downconversion provides orthogonal polarization states for the signal and idler photons, ideal for experiments requiring downstream spatial separation or polarization entanglement. Figures \ref{fig7} to \ref{fig10} display our calculations for ZGP ($o \to e + o$ or $o + e$ ) and CSP ($e \to o + e$ or $e + o$), with PPLN and PPKTP as illustrative comparisons ($o \to e + o$ or $o + e$).

\begin{figure*}[htb]\center
\includegraphics[width=1.0\textwidth]{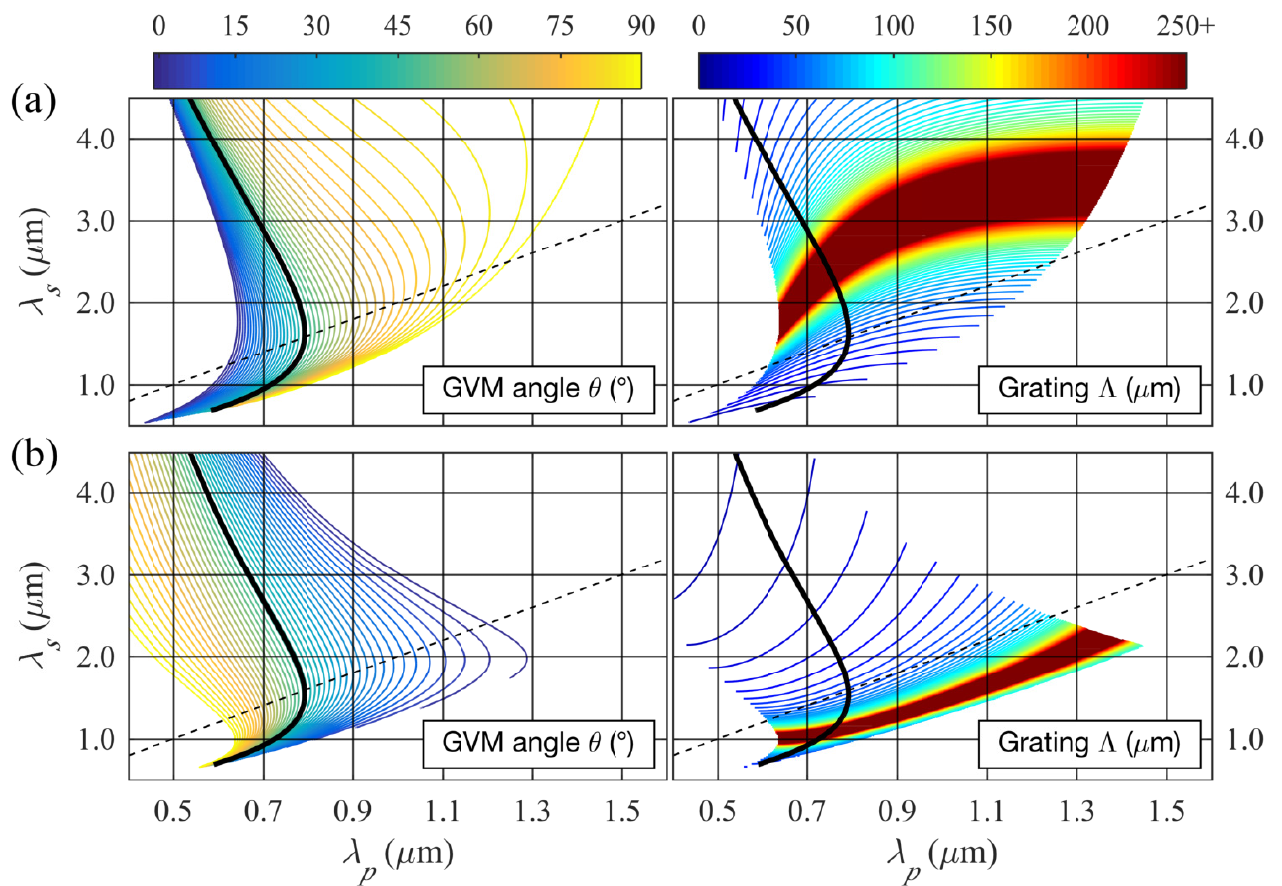}
\caption{Signal wavelength vs pump wavelength with corresponding GVM angle $\theta$ (left) and grating period $\Lambda$ (right) for $(a)$ type-II ($o \to e + o$) and $(b)$ type-II ($o \to o + e$) downconversion in PPKTP. Solid red areas in the left plot indicate grating periods longer than 250~$\mu$m, where birefringent QPM is possible.}
\label{fig7}
\end{figure*}

\begin{figure*}[htb!]\center
\includegraphics[width=1.0\textwidth]{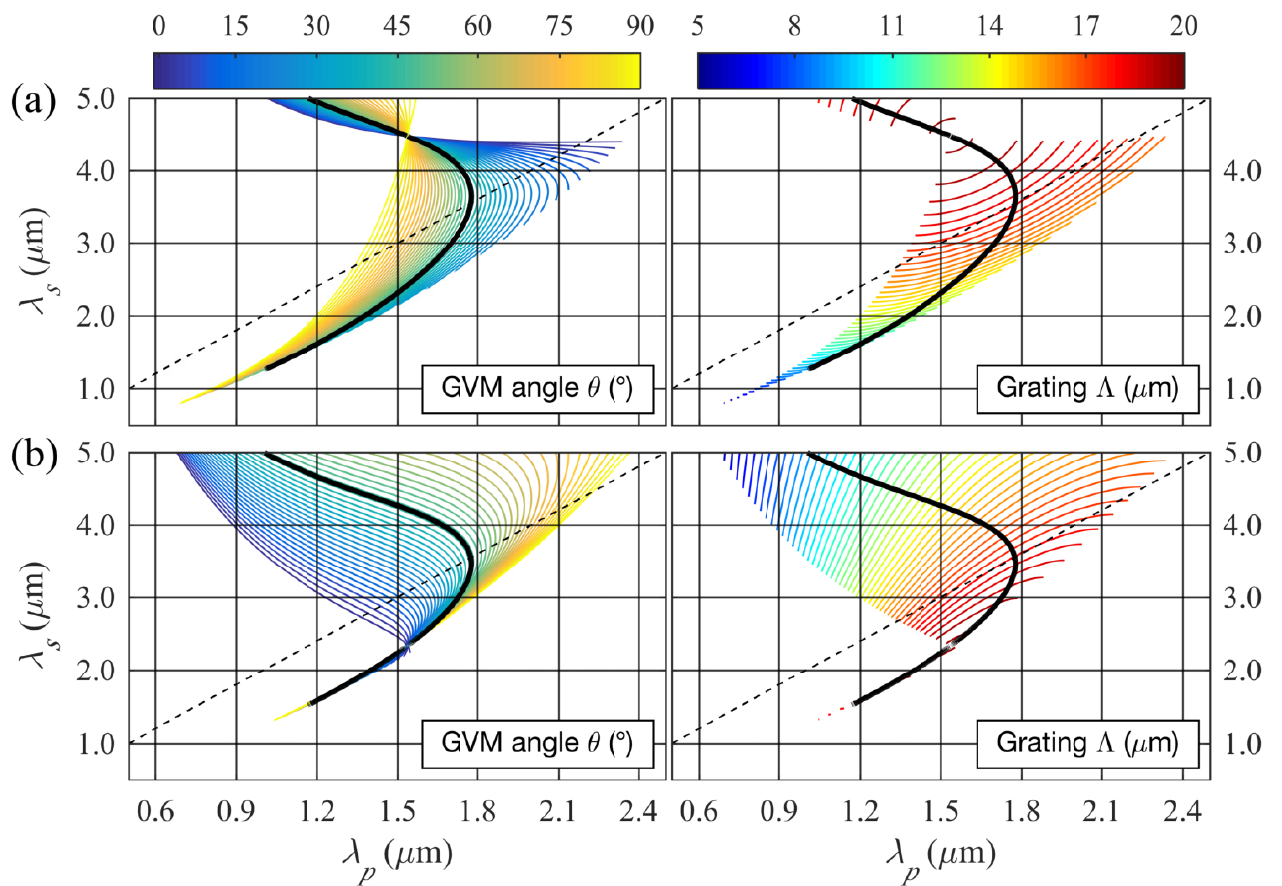}
\caption{Signal wavelength vs pump wavelength with corresponding GVM angle $\theta$ (left) and grating period $\Lambda$ (right) for $(a)$ type-II ($o \to e + o$) and $(b)$ type-II ($o \to o + e$) downconversion in PPLN.}
\label{fig8}
\end{figure*}

\begin{figure*}[htb]\center
\includegraphics[width=0.97\textwidth]{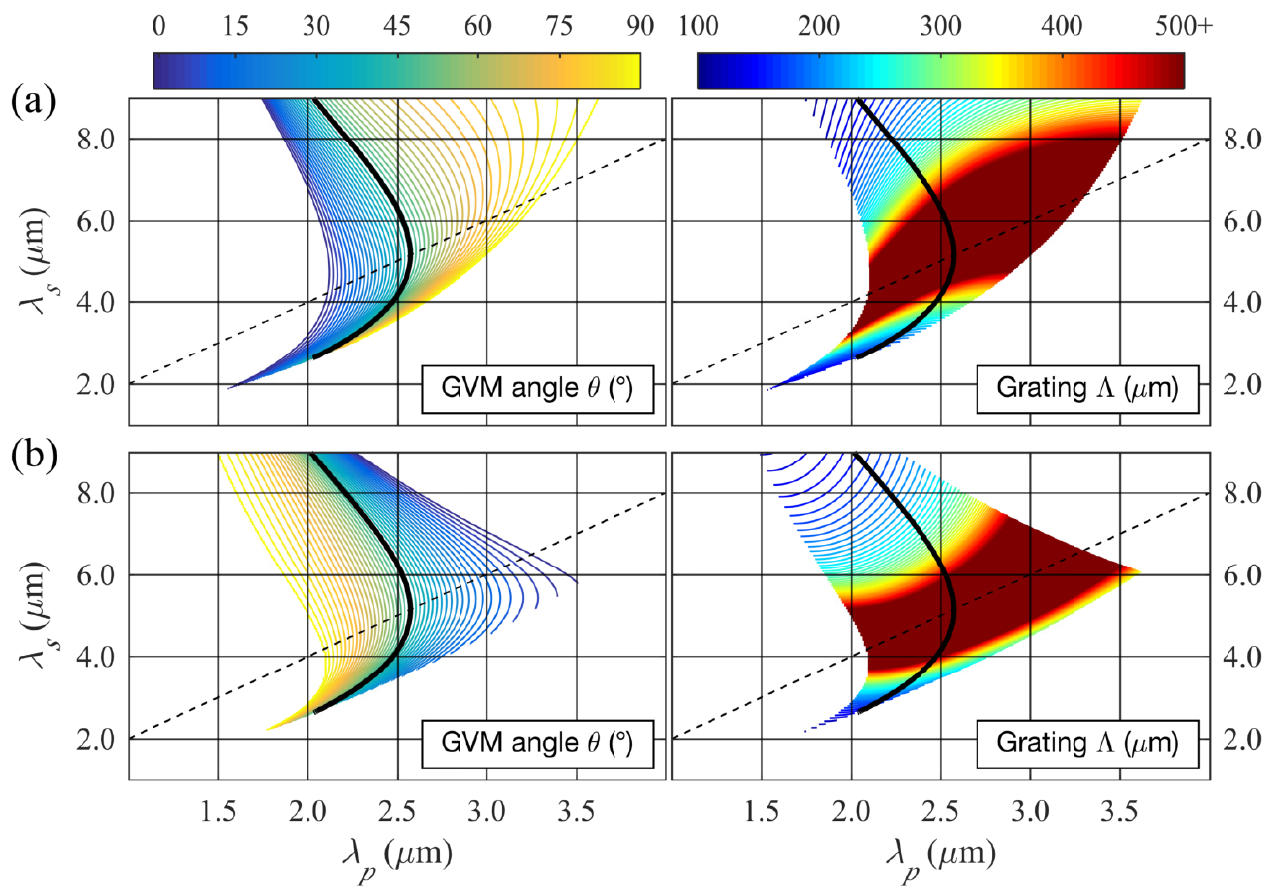}
\caption{Signal wavelength vs pump wavelength with corresponding GVM angle $\theta$ (left) and grating period $\Lambda$ (right) for $(a)$ type-II ($e \to o + e$) and $(b)$ type-II ($e \to e + o$) downconversion in CSP. Solid red areas in the left plot indicate grating periods longer than 500~$\mu$m, where birefringent QPM is possible.}
\label{fig9}
\end{figure*}

\begin{figure*}[htb!]\center
\includegraphics[width=0.97\textwidth]{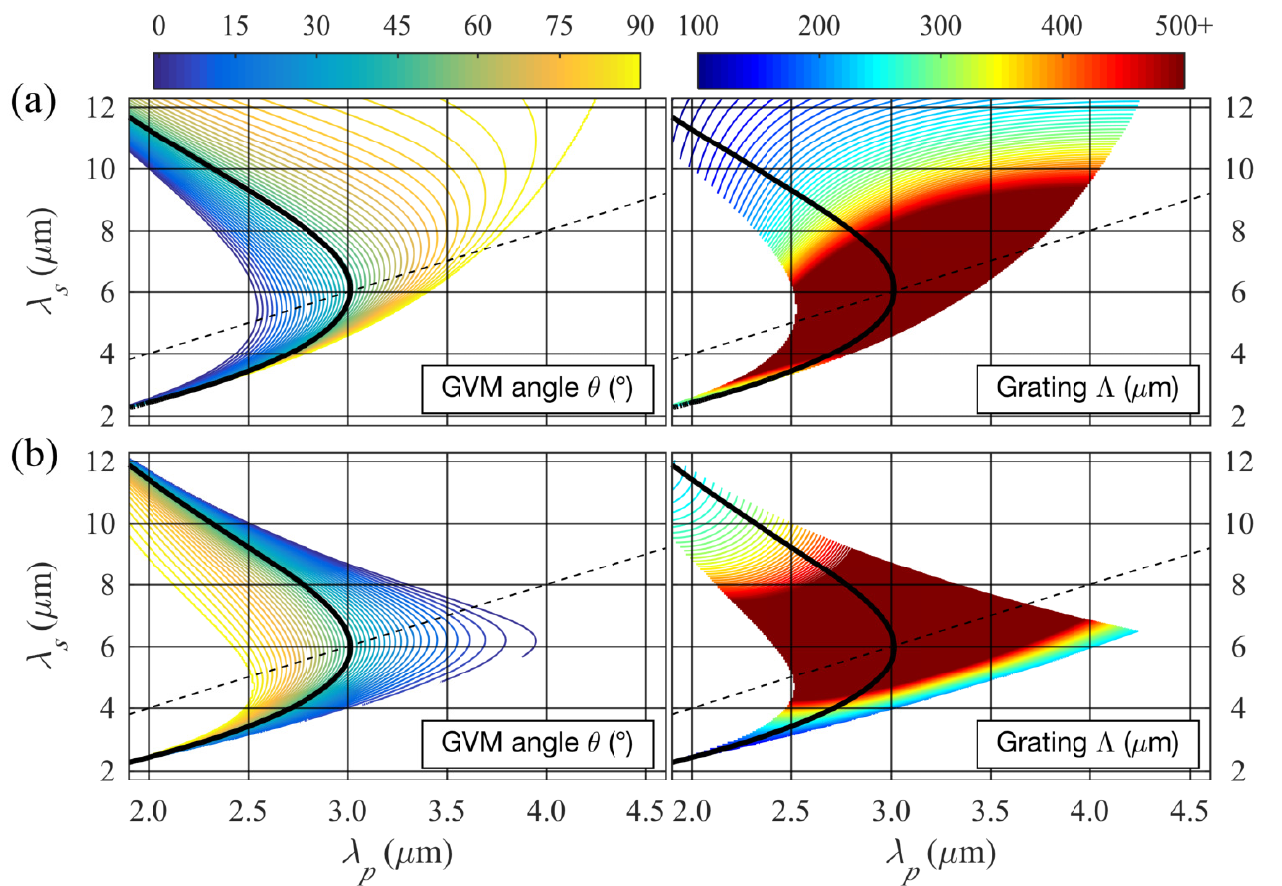}
\caption{Signal wavelength vs pump wavelength with corresponding GVM angle $\theta$ (left) and grating period $\Lambda$ (right) for $(a)$ type-II ($o \to e + o$) and $(b)$ type-II ($o \to o + e$) downconversion in ZGP. Solid red areas in the left plot indicate grating periods longer than 500~$\mu$m, where birefringent QPM is possible.}
\label{fig10}
\end{figure*}

\section{Highlighted examples of mid-IR quantum light creation}
From these numerical results it is apparent that mid-IR single photons can be generated at wavelengths of up to 13 $\mu$m in non-degenerate PDC (type-0 QPM in OP-GaAs), and up to 7.4 $\mu$m in a degenerate configuration (type-II birefringent phase-matching in ZGP). We can further identify a number of phase-matching conditions for the creation of spectrally uncorrelated, wavelength-degenerate photons. This PDC scenario is relevant for quantum-enhanced applications requiring non-classical two-photon interference, e.g. for the creation of photon-number entangled states (so-called N00N states) for phase estimation, which in the mid-IR might be of interest for non-intrusive chemical or biological sensing with a quantum advantage (see e.g. \cite{crespi2012}).

For type-II phase-matching in PPKTP, we reproduce the well-known GVM conditions from $791$~nm $\to$ $1582$~nm with $\theta=45$, for the creation of spectrally pure, indistinguishable photons in the telecom regime as e.g. implemented in \cite{Graffitti2018}. 

Moving further into the mid-IR, phase-matching is possible for degenerate PDC from $1200 \to 2400$~nm with $\theta=0$. Non-degenerate type-II PDC from $745.6 \to 1071.4 + 2451.8$~nm can be achieved in 5-mm-long birefringently phase-matched KTP with $\theta=45$. 

Type-I degenerate PDC is possible in PPLN from $784 \to 1568$~nm, providing broadband phase-matching in the telecom region. Favourable GVM conditions exist for type-II degenerate PDC from $1775$~nm $\to$ $3550$~nm with $\theta=45$, which can be achieved through OPO pumping.

The chalcopyrite crystals CSP and ZGP cannot be poled and therefore GVM conditions offering high purity and indistinguishability also require birefringent phase-matching to be of practical use. In CSP we identify degenerate type-II PDC in a 1.6-mm-long crystal from $2573 \to 5146$~nm where $\theta=45$. We also identify regions of high purity with $\theta=0$ for $2090 \to 4180$~nm ($L_{\text{crystal}}~=~1.9$ mm) and $3310 \to 6620$~nm ($L_{\text{crystal}} = 2.2$~mm), which can be pumped by holmium lasers and Yb-pumped OPOs respectively. Similarly in ZGP we find high purity degenerate type-II PDC with $\theta=0$ from $3014 \to 6028$~nm ($L_{\text{crystal}} = 8.6$~mm), along with $\theta=0$ cases from $2520 \to 5040$~nm ($L_{\text{crystal}} = 1.3$~mm) and $3692 \to 7384$~nm ($L_{\text{crystal}} = 1.5$ ~mm).

Optically isotropic semiconductors exhibit no birefringence, therefore only type-0 QPM is achievable in OP-GaP and OP-GaAs.

\section{Detection}
A key question is how to detect mid-IR single photons. In-principle suitable detectors for this spectral range can broadly be grouped into superconducting detectors---either nano-wire or transition edge sensors; narrow band-gap semiconductor photodiodes based on materials such as HgCdTe; or nonlinear up-conversion detectors which convert mid-IR single photons to wavelengths for which superconducting or semiconductor detectors are readily available.

Superconducting transition edge sensors based on tungsten have an absorption range of up to 2.5~$\mu$m \cite{Lita2008}. They offer near unity single-photon detection efficiency and photon number resolution but are quite slow and require mK cryogenic temperatures. Furthermore, the limited energy resolution of these devices restricts the wavelength range in practice to 1.5~$\mu$m \cite{Gerrits2018pc}. 

Superconducting nanowire single-photon detectors \cite{Natarajan2012} offer slightly less efficiency but are much faster and operate at less demanding temperatures of 1-2~K. SNSPDs based on WSi have been demonstrated for up to 5.5~$\mu$m detection \cite{Marsili2012,Marsili2013}, with work ongoing to extend this up to 10~$\mu$m. Challenges for SNSPDs include that the typical active area of these detectors is smaller than the mode field diameter required for mid-IR photons at 10~$\mu$m. Large-area free-space WSi SNSPDs are under development in particular for long-distance free-space communication at the single-photon level \cite{Bellei2016}. However, these options are not yet available commercially.

Semiconductor photodiodes based on InGaAs are now mature technology for 1.55~$\mu$m detection. They require cryogenic temperatures to suppress intrinsic photon-counting noise and have comparably low detection efficiency but are still widely deployed for single-photon detection in the telecommunications regime. While in principle offering a high photoresponse up to 2.6~$\mu$m, commercial options usually cover a range of no more than up to 1.7~$\mu$m. There is an ongoing effort to develop semiconductor photo-detectors based on materials with smaller band-gaps to push the detection regime into the mid-IR for applications such as space-borne astronomy or LIDAR. An early contender which has now reached some maturity is HgCdTe. A material notoriously hard to work with because of its toxicity amongst other reasons, it is now more widely available. Depending on the Cd admixture, the spectral response can be tailored to cover the entire mid-IR, from 2 to 14~$\mu$m, with the highest response at 10~$\mu$m \cite{Baker2016}. More exotic materials for which single-photon gain has been demonstrated include black arsenic phosphorus \cite{Long2017} or ultra-broadband 2D materials such as graphene \cite{Liu2014}.

Up-conversion detectors were originally developed as alternative to InGaAs detectors to improve single-photon detection characteristics at 1.55~$\mu$m \cite{Thew2006,Vandevender2007}. The infrared signal photon is mixed with a strong pump beam in a nonlinear crystal and converted to a shorter wavelength. For mid-IR single photon pairs at 3.3~$\mu$m this has recently been demonstrated in PPLN \cite{Mancinelli2017}. 

\begin{figure*}[htb]\center
\includegraphics[width=1\textwidth]{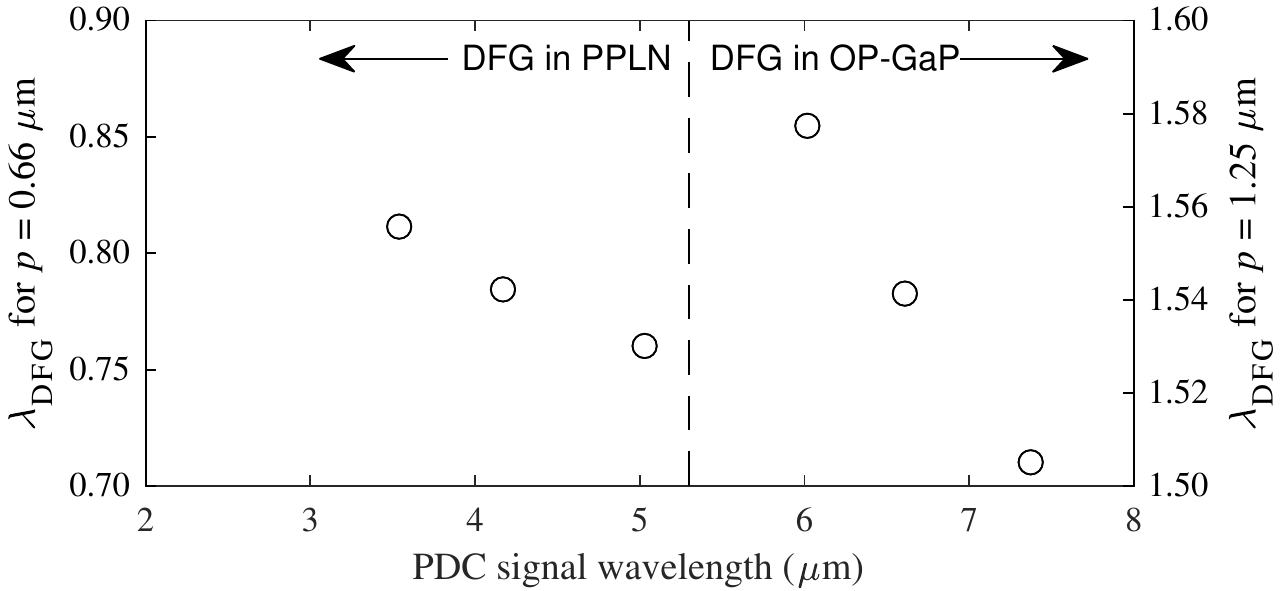}
\caption{Upconversion of mid-infrared single photons to Si and InGaAs detector bands can be achieved through type-0 DFG in PPLN or OP-GaP.}
\label{fig11}
\end{figure*}

Up-conversion detection is a natural choice in the context of this paper because the materials we discussed and the OPO pump lasers likely to be used for PDC generation would be equally suitable for the up-conversion task. For example, for the specific use cases we identified in the previous section, to produce pure single photons at 6028, 6620 and 7384~nm, upconversion to ~1550~nm could be achieved in OP-GaP with a 1250~nm seed wavelength (see Figure \ref{fig11}). Similarly, upconversion of single photons in the 3000--5000~nm band to the Si-SPAD operating region of 700--900~nm could be achieved in PPLN with a 660~nm seed wavelength. Such seed sources are available commercially or can be generated from OPOs.

Whatever option is employed, a major challenge to overcome is the thermal radiation background in the mid-IR. For example, for the mid-IR free-space SNSPD system proposed in \cite{Bellei2016}, it is estimated that even with a 1~$\mu$m bandpass filter, stray background radiation at 10~$\mu$m would amount to more than 10$^8$ noise photons per second, drowning out the signal. For up-conversion detection, it has been shown that a significant amount of broadband thermal background is converted along with the signal \cite{Barh2018}. Potential solutions to this obstacle are narrow-band filtering---although narrowband filters at the desired wavelength might not be readily available---optical gating, which for photon-pair production is standard practice anyway, and optical shuttering, which for heralded photon pairs can be done on nanosecond time-scales \cite{Brida2011}. 

Single-photon upconversion is most efficient in waveguides and therefore conventionally limited to the single-mode regime, which provides a challenge for mid-IR single-photon imaging applications such as long-range depth imaging~\cite{pawlikowska2017single}. However, a 2D imaging up-conversion system based on PPLN has been demonstrated for up to 5.5 $\mu$m \cite{Dam2012}. Alternatively, this could be addressed with photonic lanterns which adiabatically transform a multi-mode pixel array to linear single-mode arrays \cite{thomson2011ultrafast}.

\section{Conclusion}
As we have shown, recent advances in nonlinear materials allow for the creation and detection of correlated photon pairs or heralded single photons at mid-infrared wavelengths via parametric downconversion. The materials we discussed are not quite yet of the high optical quality expected for mature nonlinear crystals such as PPLN and PPKTP, however rapid progress is being made towards closing this gap. Waveguide integration is not straightforward --- ZGP and CSP do not lend themselves to refractive index modification, while OP-GaAs waveguides suffer from high propagation losses. OP-GaP waveguide structures are being actively developed, but have not been demonstrated \cite{schunemann2018pc}. Research is also going on into even more exotic crystals such as orientation-patterned zinc selenide (OP-ZnSe), barium thiogallate (BaGa$_4$S$_7$) and barium sellenogallate (BaGa$_4$Se$_7$), which might extend the accessible single-photon PDC regime beyond 13 $\mu$m. Many challenges remain to create wavelength-agile quantum light sources and the corresponding detectors, however there is also a large payoff in enabling quantum enhancements for mid-IR applications.

\begin{acknowledgments}
\noindent\textbf{Acknowledgments}

\noindent This work was supported by the Engineering and Physical Sciences Research Council (grant number EP/N002962/1). FG acknowledges studentship funding from EPSRC under grant no. EP/L015110/1. RM acknowledges support from a Heriot-Watt University Research Fellowship.
\end{acknowledgments}

\bibliography{bibliography.bib}{}
\bibliographystyle{apsrev4-1}

\end{document}